\documentstyle[emulateapj]{article}

\received{}
\accepted{}
\journalid{}{}
\articleid{}{}

\slugcomment{Accepted by the Astrophysical Journal}

\lefthead{Crawford et al.}
\righthead{Radio Pulsars in the Magellanic Clouds}

\newcommand{\approxlt}{\mbox{$\;^{<}\hspace{-0.24cm}_{\sim}\;$}}
\newcommand{\approxgt}{\mbox{$\;^{>}\hspace{-0.24cm}_{\sim}\;$}}

\begin{document}

\title{Radio Pulsars in the Magellanic Clouds}

\author{Fronefield Crawford\altaffilmark{1},  
Victoria M. Kaspi\altaffilmark{1,2,8},
Richard N. Manchester\altaffilmark{3},
Andrew G. Lyne\altaffilmark{4},
Fernando Camilo\altaffilmark{5},
and Nichi D'Amico\altaffilmark{6,7}} 

\altaffiltext{1}{Department of Physics and Center for Space Research,
Massachusetts Institute of Technology, Cambridge, MA 02139}

\altaffiltext{2}{Department of Physics, Rutherford Physics Building,
McGill University, 3600 University Street, Montreal, Quebec, H3A 2T8,
Canada}

\altaffiltext{3}{Australia Telescope National Facility, CSIRO,
P.O.~Box 76, Epping, NSW 1710, Australia}

\altaffiltext{4}{University of Manchester, Jodrell Bank Observatory,
Macclesfield, Cheshire, SK11 9DL, UK}

\altaffiltext{5}{Columbia Astrophysics Laboratory, Columbia
University, 550 West 120th Street, New York, NY 10027}

\altaffiltext{6}{Osservatorio Astronomico di Bologna, via Ranzani 1,
40127~Bologna, Italy}

\altaffiltext{7}{Istituto di Radioastronomia del CNR, via Gobetti 101,
40129 Bologna, Italy}

\altaffiltext{8}{Alfred P. Sloan Research Fellow}

\begin{abstract}
We report the results of a survey of the Small Magellanic Cloud (SMC)
for radio pulsars conducted with the 20-cm multibeam receiver of the
Parkes 64-meter telescope. This survey targeted a more complete region
of the SMC than a previous pulsar search and had an improvement in
sensitivity by a factor of about two for most pulsar periods. This
survey is much more sensitive to fast young pulsars (with $P \approxlt
100$ ms) and is the first survey of the SMC with any sensitivity to
millisecond pulsars. Two new pulsars were discovered in the survey,
one of which is located within the SMC. The number of pulsars found in
the survey is consistent with the expected number derived using
several methods.  We also report the serendipitous discovery of a new
pulsar in the 30 Doradus region of the Large Magellanic Cloud
(LMC). These discoveries bring the total number of rotation-powered
pulsars currently known in the Magellanic Clouds to eight. We have
also made refined timing measurements for the new discoveries as well
as for three previously known LMC pulsars.  The age distribution of
luminous Magellanic Cloud pulsars supports the conjecture that pulsars
younger than about 5 Myr are more luminous on average than older
pulsars.
\end{abstract}

\keywords{Magellanic Clouds --- pulsars: individual (J0057$-$7201,
J0113$-$7220, J0535$-$6935)}

\section{Introduction}

The Magellanic Clouds contain the most distant population of radio
pulsars observable with current technology. Since the distances to the
Magellanic Clouds are large, only the most luminous pulsars are
detectable. However, despite the small number of pulsars currently
known, these pulsars are among the most interesting in the pulsar
population. Of the six pulsars discovered previously, one is in an
unusual binary system and two are younger than 10 kyr. For comparison,
of over 700 Galactic pulsars cataloged prior to the commencement of
the Parkes Multibeam Pulsar Survey (see below), only one is in a
similar binary system, and only five Galactic pulsars are younger than
10 kyr.

Surveying the Magellanic Clouds for radio pulsars is difficult due to
the large surface area to be covered and their large distances. Long
integrations using a telescope with a large collecting area are
needed.  With its southern latitude, the Parkes 64-meter radio
telescope in NSW, Australia is the most appropriate telescope for
conducting such searches.

The difficulty of obtaining sufficient sensitivity has been somewhat
alleviated by the installation of the 20-cm multibeam receiver at
Parkes (\cite{swb+96}). This receiver is simultaneously sensitive to
13 distinct locations on the sky which can be interleaved in separate
pointings to give complete and efficient spatial coverage of large
regions.  The Parkes Multibeam Pulsar Survey is currently using this
receiver to search for pulsars in the Galactic plane, and that survey
has discovered 580 pulsars to date (\cite{cam00}; \cite{man00};
\cite{lcm+00}). This demonstrates that the multibeam system is capable
of finding faint pulsars and is an appropriate instrument to search
for pulsars in the Magellanic Clouds.

Only one full-scale pulsar survey of the Magellanic Clouds has been
conducted previously. McConnell et al.~(1991)\nocite{mmh+91} searched
for pulsars using Parkes at a wavelength of 50 cm.  This survey effort
was split into three phases with different observing parameters used
for each phase of the survey. Although the region surveyed in each
phase was not described, the third phase of the survey was used for
five of the seven years of the effort, and we use this phase for a
sensitivity comparison with our survey. In this phase, they used a
center frequency of 610 MHz with a total bandwidth of 60 MHz split
into 24 2.5-MHz contiguous frequency channels. Two orthogonal
polarizations were used with a sampling time of 5 ms and a 5000 s
dwell time for each pointing. The second phase of their survey had a
comparable sensitivity to the third phase, but the first phase had
significantly reduced sensitivity to fast pulsars due to the much
larger sampling time of 40 ms used.

In their data reduction procedure, McConnell et
al.~(1991)\nocite{mmh+91} assumed a dispersion measure (DM) of 100 pc
cm$^{-3}$ in an initial sub-band dedispersion in their
analysis. Magellanic Cloud pulsars found to date have DMs which are
roughly centered on this value, but the observed DM range is from
about 70 pc cm$^{-3}$ to 150 pc cm$^{-3}$. Thus, their sensitivity to
pulsars with DMs far from 100 pc cm$^{-3}$ was significantly reduced.
McConnell et al.~(1991)\nocite{mmh+91} also targeted the optical bar
and the southern and western parts of the Small Magellanic Cloud (SMC)
(see their Figure 1). Our survey had more complete coverage in the
northern and eastern regions of the SMC.  McConnell et
al.~(1991)\nocite{mmh+91} reported the discovery of four pulsars,
three in the Large Magellanic Cloud (LMC) and one in the SMC.

We report the results of a more sensitive and complete survey for
pulsars in the SMC in which we have discovered one new SMC pulsar and
one foreground pulsar. Below we describe the observations and data
reduction procedure for our survey and the subsequent timing
observations. We also report on the serendipitous discovery of a
pulsar in the 30 Doradus region of the LMC, as well as on timing
observations of three previously known LMC pulsars.  We discuss our
results and summarize the currently known Magellanic Cloud pulsar
population. The two new Magellanic Cloud pulsars found in our survey,
combined with the four pulsars found by McConnell et
al. (1991)\nocite{mmh+91} and the two known X-ray pulsars in the LMC,
bring the total number of known rotation-powered pulsars in the
Magellanic Clouds to eight.

\section{Observations and Data Reduction}

\subsection{Survey Observations and Data Reduction} 

The multibeam receiver is capable of simultaneously observing 13
separate regions on the sky and has been designed to interleave
pointings in such a way that clusters can be formed from each set of
four pointings, thereby offering efficient and complete sky
coverage. System and observing details can be found elsewhere in
descriptions of the Parkes Multibeam Pulsar Survey (\cite{lcm+00};
\cite{ckl+00}; \cite{mlc+00a}). The receiver for each beam is a
dual-channel cryogenically-cooled system sensitive to orthogonal
linear polarizations.  A 288-MHz bandwidth is centered on a frequency
of 1374 MHz. Each beam has a full-width half-power diameter of $\sim$
14$'$ with an average system noise temperature of 21 K (33 Jy) on cold sky. A
filterbank system covering the 288-MHz bandwidth which has 96
separate 3-MHz channels for each polarization for each beam was used.

\begin{figure}
\plotone{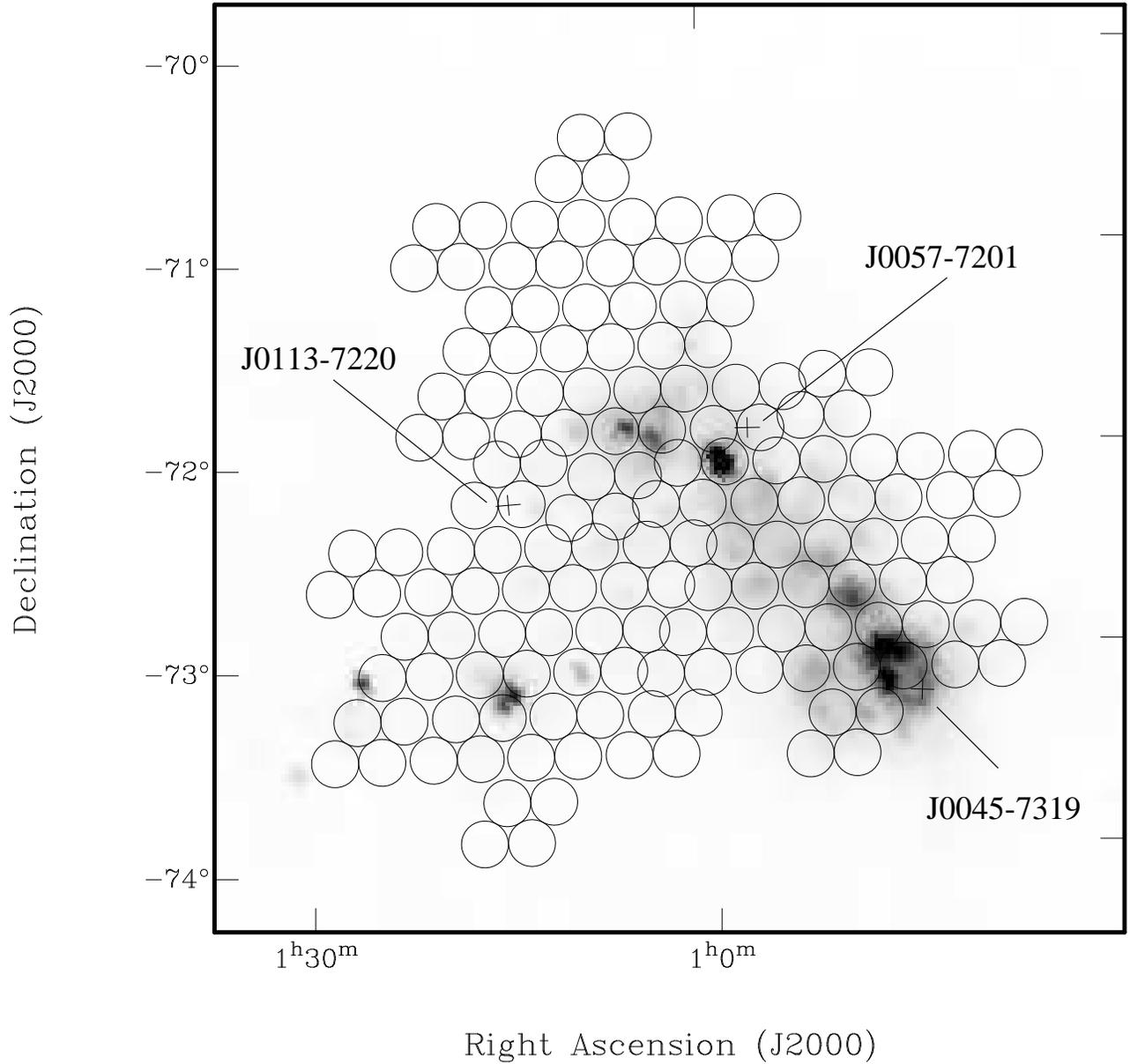}
\caption{Multibeam survey coverage of the SMC. Each circle represents
the half-power beamwidth of a single beam of the receiver. A total of
12 pointings (156 beams) were interleaved for tiled coverage.  IRAS
60-$\mu$m emission is indicated in grayscale and roughly defines
our survey region. The locations of the three pulsars discovered in
the direction of the SMC to date are indicated by crosses. PSR
J0057$-$7201 is a foreground pulsar while the other two lie in the
SMC.}
\label{fig-1}
\end{figure}

During observing, detected signals from each channel were square-law
detected and added in polarization pairs before undergoing high-pass
filtering. The signals were then one-bit digitized every 0.25 ms and
recorded on magnetic tape for processing and analysis. We
observed each of a total of 12 pointings (156 separate beams) for 8400
s, providing a limiting flux density sensitivity of $\sim$ 0.08 mJy
for most of the pulsar period range.  The total area surveyed
was $\sim$ 6.7 sq. deg., corresponding to an area of $\sim$ 6.6
kpc$^{2}$ for an assumed distance to the SMC of 57 kpc (\cite{col98}).
Figure \ref{fig-1} illustrates the distribution of beams in our survey
overlaid on an IRAS 60-$\mu$m
image\footnote{http://www.ipac.caltech.edu.} of the emission in the
region in grayscale.

Off-line processing was conducted on Sun workstations. Each beam was
dedispersed for 191 trial values of DM varying from 0 to 442 pc
cm$^{-3}$, a much greater range than that of the DMs of the known
Magellanic Cloud pulsars. The dedispersed data for each DM trial were
high-pass filtered before an amplitude spectrum was formed using a
2$^{25}$-point FFT. Portions of the spectrum (typically a few percent)
were masked for each DM trial according to whether radio-frequency
interference appeared significantly in the zero-DM spectrum. Higher
sensitivity to narrow pulses was achieved through incoherent addition
of up to 16 harmonics (including the fundamental).

The strongest candidates in each DM trial were recorded if they had a
signal-to-noise ratio greater than 7 and appeared in at least 6 of the
DM trials (in order to avoid an excess of spurious candidates). The
original data were then reprocessed by dedispersing and folding the
data over a small range of DMs and periods centered on the candidate
values. Follow-up confirmation observations were undertaken in the
half-dozen cases where the resulting folded data showed a pulsar-like
signature (i.e., in cases where a signal was well-localized in
period/DM phase space).

The relatively high frequency of the multibeam receiver compared to
the lower frequency used in the McConnell et al.~(1991)\nocite{mmh+91}
survey is an advantage since dispersive smearing and interstellar
scattering are greatly reduced at this frequency, thereby allowing the
detection of fast, distant, and highly scattered pulsars which may
have been previously missed.

\begin{figure}
\plotone{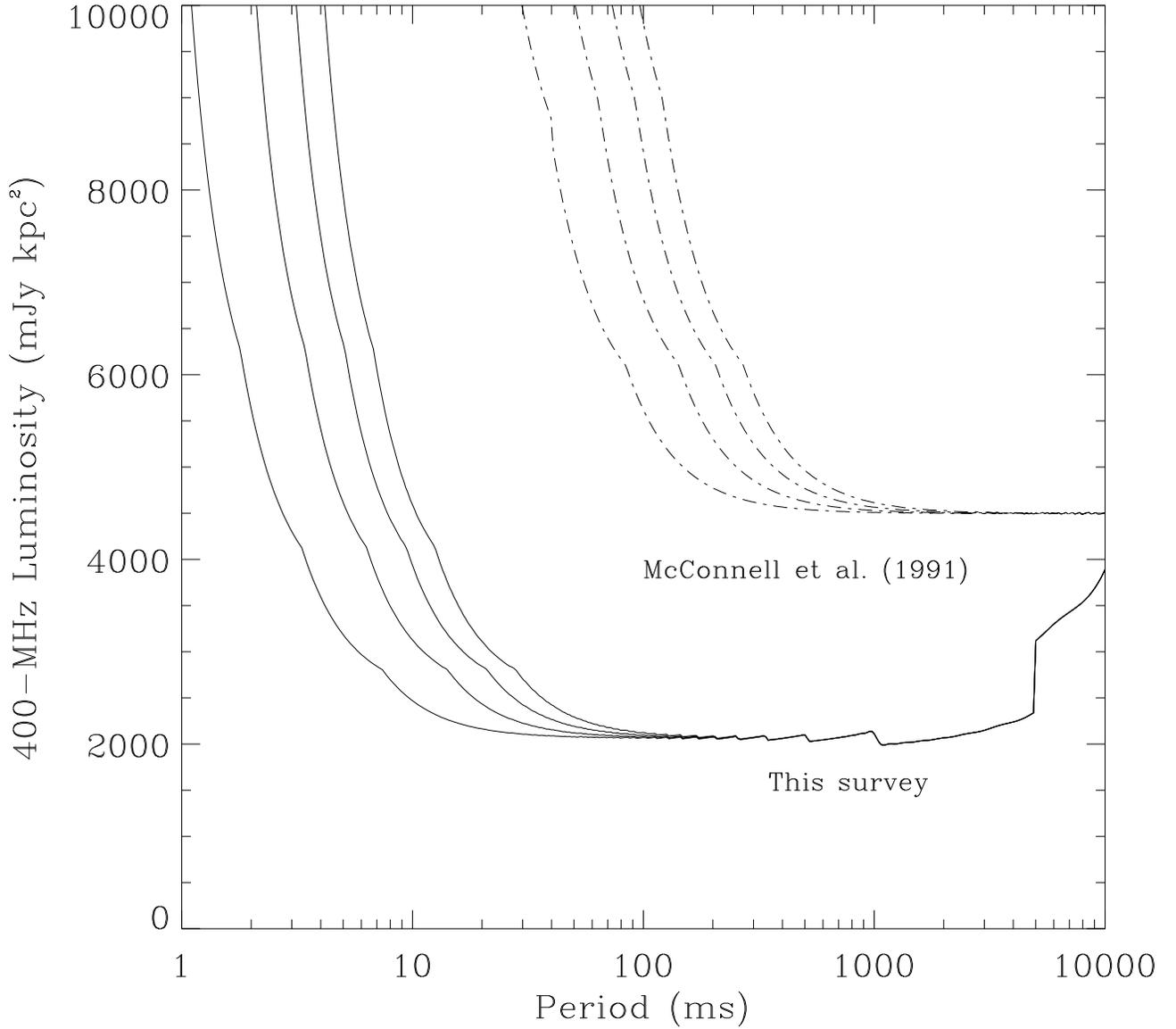}
\caption{Limiting luminosity of pulsars in the SMC as a
function of pulsar period for our survey (solid lines) and for 
the third phase of the
McConnell et al.~(1991) survey (dashed-dotted lines). A 5\% pulsed duty
cycle is assumed in each case. The two surveys were conducted at
different radio frequencies, so the sensitivity has been scaled to a
minimum detectable 400-MHz luminosity assuming a standard pulsar
spectral index ($\alpha = -1.6$) and an SMC distance of 57 kpc. Curves
for DM values of 50, 100, 150, and 200 pc cm$^{-3}$ are shown (from
left to right).  The detailed
sensitivity calculation used here follows that found in Crawford
(2000)
for the Parkes Multibeam Pulsar Survey.}
\label{fig-2}
\end{figure}

Since our survey was conducted at a different frequency than the
McConnell et al.~(1991)\nocite{mmh+91} survey, we must scale the
sensitivity of each survey in order to compare them. We have scaled
each survey to a 400-MHz luminosity ($L_{400}$), assuming a distance
to the SMC of 57 kpc (\cite{col98}) and a typical pulsar spectral
index $\alpha$ = $-$1.6 in each case (\cite{lylg95}), where $\alpha$
is defined according to $S \sim \nu^{\alpha}$. Figure \ref{fig-2}
shows a comparison of the sensitivity of the two surveys for a variety
of pulsar periods and DMs. Our survey is significantly more sensitive
to fast young pulsars (with $P \approxlt 100$ ms) and is the first
survey of the SMC which has any sensitivity to millisecond pulsars.

\subsection{Timing Observations and Data Reduction}

New pulsars discovered in our survey were timed regularly for about a
year following their discovery. The observing system used for timing
these pulsars at Parkes is identical to that used in the survey with
the exception that the data were only recorded from the center beam of
the multibeam receiver. Several additional timing observations were
made at 70 cm. The start time of each observation was recorded and was
synchronized with an observatory time standard, and data were usually
recorded on magnetic tape. After dedispersion, the resulting time
series was then folded at the topocentric rotation period of the
pulsar, generating a single pulse profile for each observation. A
topocentric pulse time-of-arrival (TOA) was obtained for each timing
observation by cross-correlating the pulse profile with a high
signal-to-noise template profile. Spin and astrometric parameters were
then determined using the TEMPO software
package,\footnote{http://pulsar.princeton.edu/tempo.} and the JPL
DE200 planetary ephemeris (\cite{sta90}). The TEMPO package converts
each TOA to the solar system barycenter and refines the estimated spin
and astrometric parameters by minimizing the residual differences
between measured and model TOAs over the span of observations. Timing
data for three known pulsars in the LMC were previously obtained at
several observing frequencies at Parkes with an observing system
described elsewhere (\cite{kas94}; \cite{jml+95}). These data were
reprocessed to obtain refined timing parameters.

\section{Results}

We have discovered two new pulsars in our survey of the SMC. One pulsar,
PSR J0057$-$7201, has a significantly smaller DM than the population
of known Magellanic Cloud pulsars, and therefore we believe it is a
foreground Galactic object. The other new pulsar, PSR J0113$-$7220,
has a DM of 125 pc cm$^{-3}$, a value which is somewhat larger than
the only previously discovered SMC pulsar, PSR J0045$-$7319,
indicating that it is located within the SMC.

As part of an ongoing campaign to search for radio pulsations from
X-ray targets in the LMC, while pointing at the X-ray pulsar PSR
J0537$-$6910 (\cite{mgz+98}; \cite{ckm+98a}), we have serendipitously
discovered one new pulsar, PSR J0535$-$6935, in one of the outlying
beams of the multibeam receiver. With a DM of 89 pc cm$^{-3}$, this
pulsar almost certainly lies in the LMC, making it the sixth LMC
rotation-powered pulsar known.

\begin{figure}
\plotone{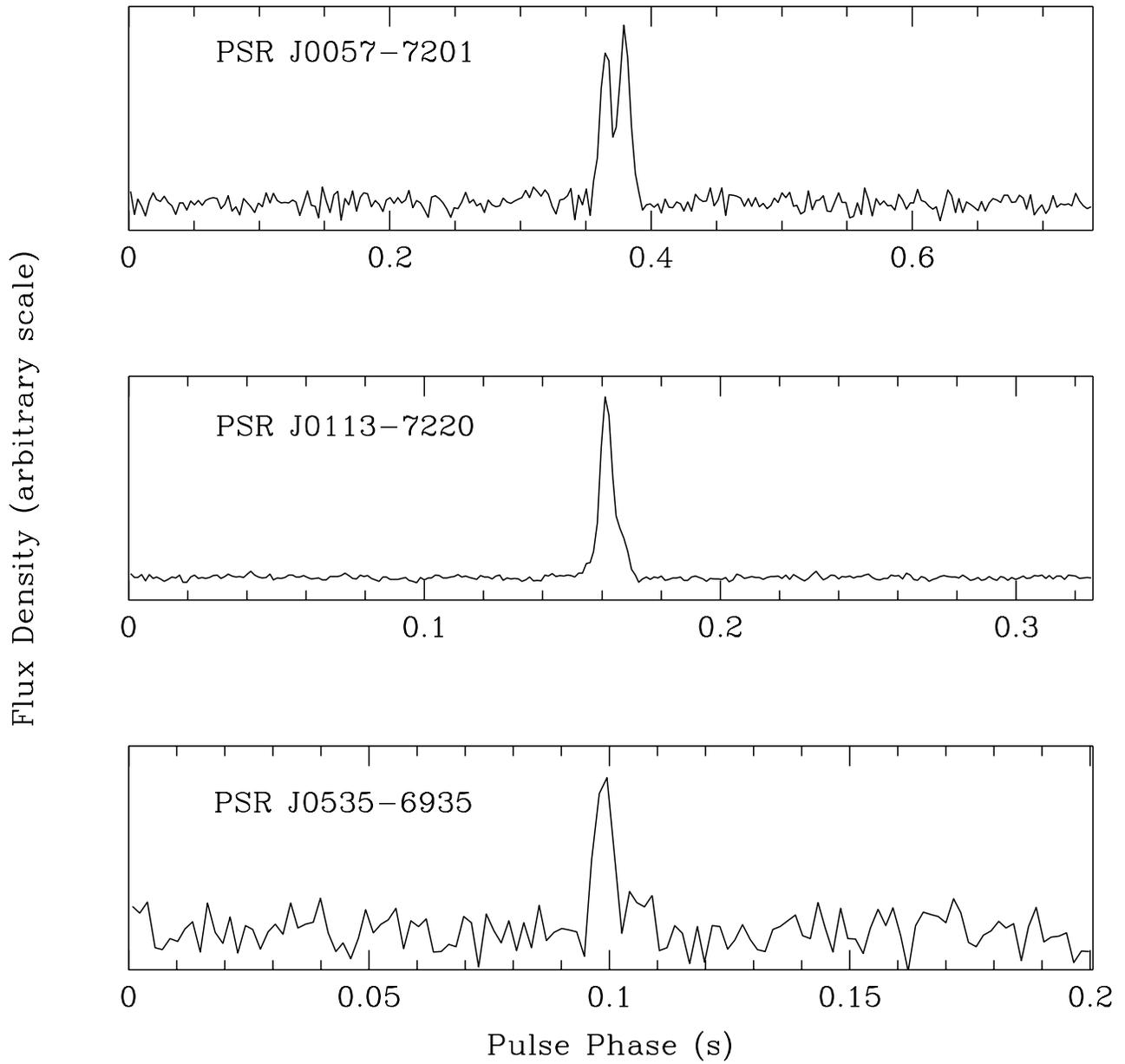} 
\caption{20-cm integrated total intensity profiles for PSRs
J0057$-$7201, J0113$-$7220, and J0535$-$6935, the three newly
discovered radio pulsars reported here.}
\label{fig-3}
\end{figure}

The measured parameters for the three newly discovered pulsars are
given in Table \ref{tbl-1}, and integrated 20-cm profiles are given in
Figure \ref{fig-3}. Flux densities at 1374 MHz were estimated for
these pulsars from the detection strengths in the discovery
observations. The results of timing measurements of the previously
known pulsars are given in Table \ref{tbl-2}. Table \ref{tbl-4new}
summarizes the general characteristics of the eight pulsars now known
in the Magellanic Clouds, and we briefly describe each of these in
turn below, as well as the new Galactic pulsar PSR J0057$-$7201.

\subsection{PSR J0045$-$7319}

\begin{figure}
\plotone{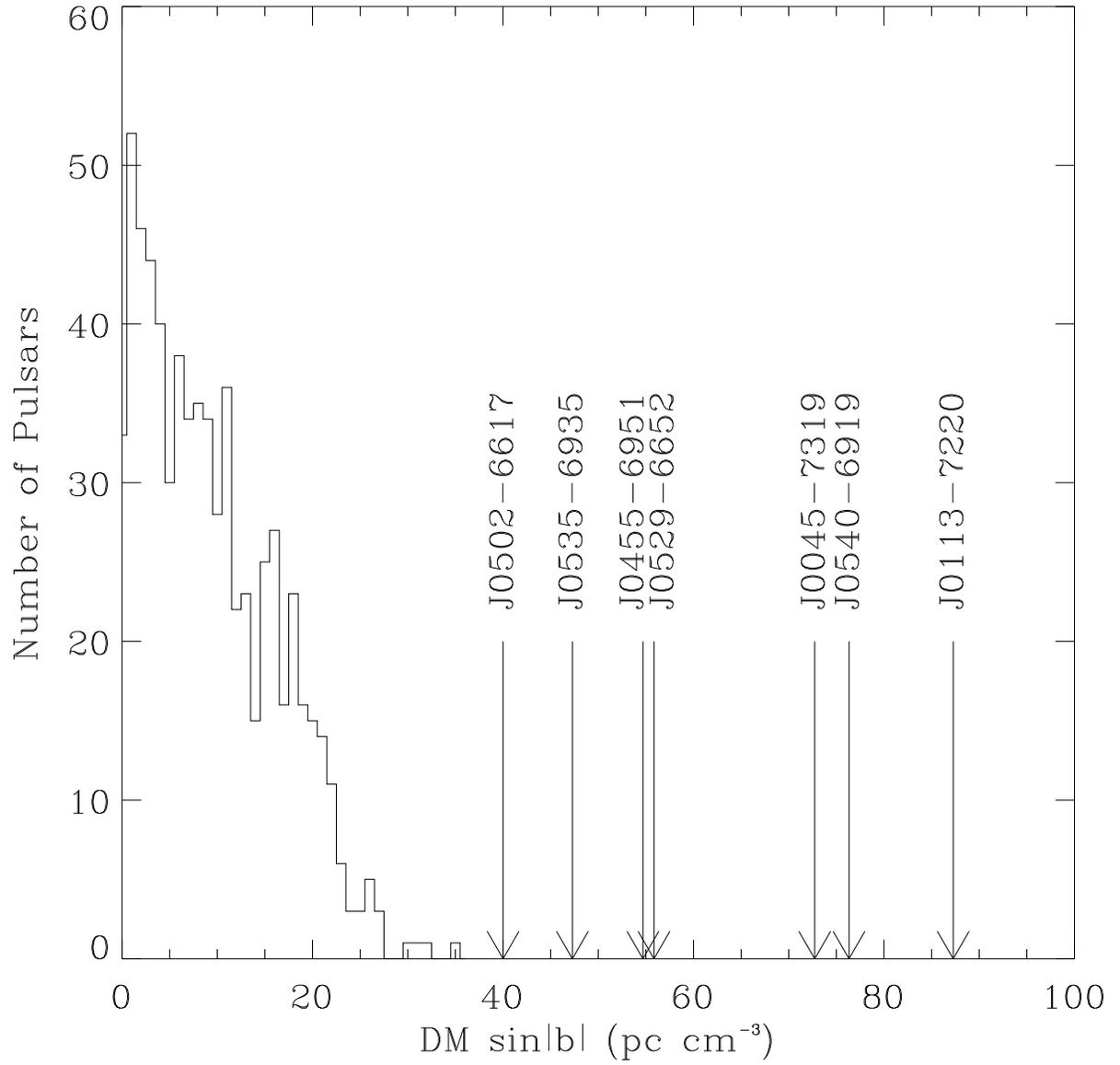}
\caption{Histogram of ${\rm DM}$~$\sin |b|$ for 681 Galactic pulsars.
Values of ${\rm DM}$~$\sin |b|$ for the seven Magellanic Cloud pulsars
for which the DM has been measured are indicated by arrows.}
\label{fig-4}
\end{figure}

PSR J0045$-$7319 was discovered in the SMC in the previous survey of
the region by McConnell et al.~(1991)\nocite{mmh+91}. Timing
observations subsequently showed it to be in a 51-day binary orbit
around a B1 class V star with a mass of 9 $M_{\odot}$ (\cite{kjb+94}).
Of over 700 pulsars cataloged by Taylor, Manchester, \& Lyne
(1993)\nocite{tml93}\footnote{The Public Pulsar Catalogue 
containing entries
for 772 pulsars can be found at http://www.atnf.csiro.au/research/pulsar/psr.} prior to
the start of the Parkes Multibeam Pulsar Survey, only one other
pulsar, PSR B1259$-$63, has been shown to be in a similar system
(\cite{jml+92}). The large ${\rm DM}$~$\sin |b|$ of PSR J0045$-$7319
(Figure \ref{fig-4}) suggests that it is in the SMC. Its location in
the SMC was confirmed by its association with the B star companion,
which was known to be in the SMC.

We detected PSR J0045$-$7319 in several diagnostic observations in the
survey. We notice an unusual broadband ($\approxgt 300$ MHz) effect at
20 cm in which large amplitude fluctuations occur on time-scales of
tens of minutes. 
In some timing observations, the pulsar is not detectable at all at 20
cm. It is possible that this behavior is an unusual form of refractive
scintillation, perhaps related to the binary nature of the pulsar, but
a detailed study of this behavior is beyond the scope of this paper.

\subsection{PSR J0057$-$7201}

PSR J0057$-$7201 is a 738-ms pulsar that was discovered in our
survey. From its DM of 27 pc cm$^{-3}$, which is significantly smaller
than the lowest known DM for a suspected Magellanic Cloud pulsar (69
pc cm$^{-3}$ for PSR J0502$-$6617), we conclude that it is a
foreground object. The Taylor \& Cordes~(1993)\nocite{tc93}
DM-distance model indicates that the distance to PSR J0057$-$7201 is
greater than 2.5 kpc, which is the limit of the Galactic plasma in the
direction of this pulsar.  However, the value of ${\rm DM}$~$\sin |b|$
for PSR J0057$-$7201 is 19 pc cm$^{-3}$, still well within the
distribution of Galactic pulsars (Figure \ref{fig-4}). Thus we cannot
conclude that it is located within the SMC. The pulsar exhibits
significant scintillation on time-scales comparable to the length of
the timing observations ($\sim$ 1 h) and was only detectable in about
half of the 20-cm timing observations and in none of the 70-cm
observations.

\subsection{PSR J0113$-$7220}

PSR J0113$-$7220 was first discovered in our SMC survey and has a
326-ms period. The pulsar shows no noticeable scintillation,
consistent with the known population of luminous LMC pulsars.  Its DM
of 125 pc cm$^{-3}$ is larger than that of PSR J0045$-$7319, the only
other known SMC pulsar, and implies that it is also located within the
SMC.  Radio timing results for PSR J0113$-$7220 (Table \ref{tbl-1})
reveal a characteristic age of $\sim$ 1 Myr. PSR J0113$-$7220 is also
very luminous with a narrow profile peak. It is the most luminous
pulsar currently known in either of the Magellanic Clouds.

\subsection{PSR J0535$-$6935}

\begin{figure}
\plotone{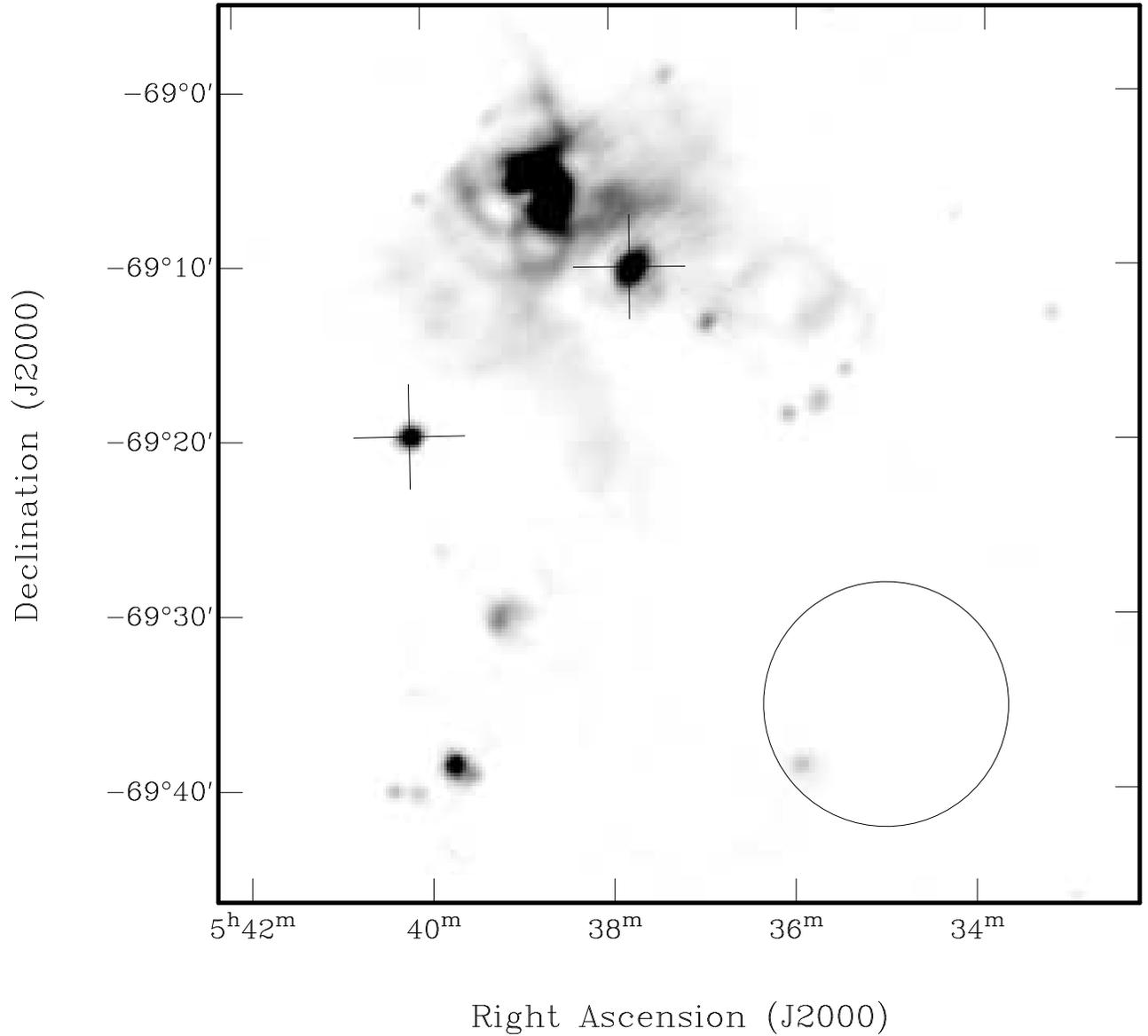}
\caption{A Molonglo Observatory Synthesis Telescope 843 MHz image
of the 30 Doradus region of the LMC. The two crosses indicate
the positions of two young pulsars, PSR J0537$-$6910 (upper right) and
PSR J0540$-$6919 (lower left), which are associated with 
SNR 0538$-$69.1 (N157B) and SNR 0540$-$69.3 respectively. The discovery of PSR
J0535$-$6935 in an outlying beam of the multibeam receiver during
a search for radio pulsations from PSR J0537$-$6910 with the center
beam is reported here; the error circle for this pulsar's location is
shown. The faint source within the circle is too bright to be the
pulsar.}
\label{fig-6}
\end{figure}

PSR J0535$-$6935 was discovered serendipitously in the 30 Doradus
region of the LMC in one of the outlying beams of the multibeam
receiver during a deep search for radio pulsations from PSR
J0537$-$6910 with the center beam (\cite{ckm+98a}). Figure \ref{fig-6}
shows an 843-MHz radio image of the 30 Doradus region from the Sydney
University Molonglo Sky
Survey\footnote{http://www.astrop.physics.usyd.edu.au/SUMSS.}, with
the locations of the three known pulsars in that region indicated. PSR
J0535$-$6935 proved too faint for regular timing, so an estimate of
$\dot{P}$ was made by comparing the barycentric period in observations
obtained one year apart. Table \ref{tbl-1} lists the result. The
positional uncertainty remains large ($\sim$ 7$^{\prime}$, the radius
of the detection beam). The pulsar was detected in observations of
length 21600 s and 17200 s and would have been too faint to detect in
our standard 8400 s integrations in the SMC survey.

\subsection{PSR J0537$-$6910}

PSR J0537$-$6910 is a 16-ms X-ray pulsar in the 30 Doradus region of
the LMC (Figure \ref{fig-6}). This Crab-like pulsar was first detected
in X-rays (\cite{mgz+98}) and is associated with the plerionic
supernova remnant (SNR) 0538$-$69.1 (N157B) in the LMC. The pulsar has
a characteristic age of $\sim$ 5 kyr, making it the fastest young
rotation-powered pulsar currently known and one of the few pulsars
with a confirmed SNR association.  Efforts to find a radio counterpart
to this X-ray pulsar have so far been unsuccessful at both 50 cm and
20 cm (\cite{ckm+98a}), but constraints on the radio luminosity are
poor due to the large distance.

\subsection{PSR J0540$-$6919 (B0540$-$69)}

PSR J0540$-$6919 (B0540$-$69) is a 50-ms Crab-like pulsar in the 30
Doradus region of the LMC (Figure \ref{fig-6}) which was first
discovered in X-rays (\cite{shh84}) and subsequently detected at radio
wavelengths (\cite{mml+93}). The pulsar is associated with the composite 
SNR 0540$-$69.3 in the LMC and is very young, with a characteristic
age of 1.7 kyr, making this system a twin of the Crab system.  The
measured DM of 146 pc cm$^{-3}$ for PSR J0540$-$6919 is the largest
yet measured for any pulsar in the Magellanic Clouds.  An electron
density of $n_{e} \sim 2$ cm$^{-3}$ has been estimated for the 30
Doradus region in the direction of SNR 0538$-$69.1 (N157B) (Lazendic
et al. 2000)\nocite{ldh+00}. The excess DM of $\sim 50$ pc cm$^{-3}$
for PSR J0540$-$6919 relative to other known LMC pulsars would suggest
a characteristic size of $\sim$ 25 pc for a region with a similarly
enhanced plasma density.  This is significantly larger than the $\sim$
1 pc size of SNR 0540$-$69.3 ($\sim 0.1'$ at 50 kpc) and suggests that
the larger DM of PSR J0540$-$6919 could be accounted for by a region
of enhanced plasma density extending well beyond SNR 0540$-$69.3
itself.

\subsection{PSR J0455$-$6951, PSR J0502$-$6617, and PSR J0529$-$6652}

These three pulsars were first discovered in the 50-cm survey of the
Magellanic Clouds by McConnell et al.~(1991)\nocite{mmh+91} but were
not subsequently timed by them.  Timing results were first obtained by
Kaspi (1994)\nocite{kas94} but not published elsewhere. In Table
\ref{tbl-2} we present updated and refined timing results for these
pulsars which we obtained by reprocessing the timing data.

\section{Discussion} 

\subsection{Dispersion Measures} 

The Taylor \& Cordes~(1993)\nocite{tc93} model of the electron
distribution in the Galaxy incorporates layers of electrons in which
the density decreases with increasing $z$-distance from the Galactic
plane. This clearly limits the Galactic contribution to the DM as a
function of Galactic latitude. One reason that pulsars are believed to
be in the Magellanic Clouds is their large DMs, which exceed those
expected for Galactic pulsars at that Galactic latitude. The quantity
${\rm DM}$~$\sin |b|$ is a measure of the $z$-contribution to the DM
and can be used to distinguish Galactic from extragalactic pulsars.
Figure \ref{fig-4} shows a histogram of the measured $\mbox{DM}$~$\sin
|b|$ values for 681 Galactic pulsars. There is a clear dropoff close
to zero at about 27 pc cm$^{-3}$ with several pulsars between 30 pc
cm$^{-3}$ and 35 pc cm$^{-3}$. The ${\rm DM}$~$\sin |b|$ values of
Magellanic Cloud pulsars range from 40 pc cm$^{-3}$ to about 90 pc
cm$^{-3}$, a much larger range than the Galactic population
itself. One reason for this is that the projected $z$-contribution of
electrons from the Magellanic Clouds is not a flat disk (like our
Galaxy), but rather is extended in the line of sight.
 
One interesting question to ask is whether the spread in DM seen in
the known pulsar population can reveal anything about the Magellanic
Clouds themselves. The observed range for the two SMC pulsars is 105
pc cm$^{-3}$ to 125 pc cm$^{-3}$.  The observed range in the LMC
(where there are more known pulsars) is about 80 pc cm$^{-3}$, from
the smallest value of about 70 pc cm$^{-3}$ to the largest value of
about 150 pc cm$^{-3}$.  Assuming an electron density of 0.03
cm$^{-3}$ for the Magellanic Clouds which is comparable to estimated
mean Galactic values (\cite{mt77}; \cite{spi78}; \cite{tc93}), the DM
spread corresponds to a distance range of 0.7 kpc and 2.7 kpc for the
SMC and LMC respectively.

From distance measurements of 161 Cepheids in the SMC,
Mathewson~(1985)\nocite{mat85} has estimated that the depth of the SMC
is between 20 kpc and 30 kpc, much larger than the projected size of
the SMC on the sky of $\sim$ 4 kpc ($\sim$ 4$^{\circ}$). The observed
DM spread for pulsars in both Magellanic Clouds indicates
line-of-sight depths that are much smaller than this, as indicated
above. In fact, for an SMC depth of 20-30 kpc, the implied mean
electron density from the pulsar DM distribution (assuming that
the pulsars are separated by a significant fraction of the SMC size) 
would be $n_{e} \approxlt 
0.001$ cm$^{-3}$, much smaller than Galactic values.  Our survey is
sensitive to high-DM pulsars and our data processing includes DM
$\approxlt$ 450 pc cm$^{-3}$, but we did not detect any pulsars with
DM $>$ 125 pc cm$^{-3}$. Our results are more consistent with the
suggestion of Zaritsky et al.~(2000)\nocite{zhg+00} that the inner
part of the SMC is roughly spherical in morphology, though a
comparison of the older population of Cepheids with the younger
population of pulsars may not account for systematic differences in
their distributions due to their ages. Should a significant number of
additional SMC pulsars be found in the future in this same DM range,
this will be at odds with an elongated line-of-sight morphology for
the SMC.

\subsection{Expected Number of Detectable Pulsars in the Magellanic Clouds}

Establishing formation rates of neutron stars and determining the
beaming fraction and luminosity characteristics of pulsars is
important for understanding the birth and emission characteristics of
the pulsar population. Although the Galactic pulsar population
currently provides a much larger sample of pulsars with which to model
these parameters, it suffers from selection effects and distance
uncertainties. The Magellanic Clouds suffer much less from these
effects.  Here we compare the observed number of pulsars in the SMC
with the number predicted from several methods using different model
assumptions.

We follow the method of McConnell et al.~(1991)\nocite{mmh+91}
to estimate the number of SMC pulsars expected to be detectable in our
survey. The number of potentially observable pulsars in the SMC,
$N_{\rm SMC}$, can be estimated using the number of potentially
observable pulsars in our solar neighborhood which is defined as a
cylinder with a base area of 1 kpc$^{2}$ in the Galactic disk. We
scale this number by a factor which includes the mass ratio of the SMC
and the solar neighborhood and their relative star formation rates,

\begin{equation}
N_{\rm SMC} = R \frac{N_{\rm SN} M_{\rm SMC}}{M_{\rm SN}} .
\label{eqn:nsmc}
\end{equation}

\noindent 
Here $R$ is the star formation rate in the SMC relative to the solar
neighborhood, $N_{\rm SN}$ is the number of potentially observable
pulsars in the solar neighborhood, and $M_{\rm SMC}$ and
$M_{\rm SN}$ are the masses of the SMC and solar neighborhood
respectively.

Lequeux (1984)\nocite{leq84} indicates that the star formation rate
per unit total mass in the SMC is greater than the solar neighborhood,
yielding $R$ = 1.6.  The total masses of the solar neighborhood and
SMC are estimated to be $9 \times 10^{7} M_{\odot}$ and $1.8 \times
10^{9} M_{\odot}$, respectively (\cite{vlm+80}). The number of
potentially observable pulsars with $L_{400} > 1$ mJy kpc$^{2}$ in the
solar neighborhood has been estimated by Lyne et
al.~(1998)\nocite{lml+98}. They derive a local space density of 30
$\pm$ 6 kpc$^{-2}$ and 28 $\pm$ 12 kpc$^{-2}$ for non-recycled and
millisecond pulsars respectively above this luminosity limit. These
estimates include beaming effects and hence are a fraction of the true
underlying pulsar population.  Putting these numbers in Equation
\ref{eqn:nsmc} gives 1000 $\pm$ 200 and 900 $\pm$ 400 for the number of
potentially observable non-recycled and millisecond pulsars
respectively in the SMC with $L_{400} > 1$ mJy kpc$^{2}$.

We can convert this result into an estimate of the number of pulsars
detectable in our survey, which is sensitive to pulsars with $L_{400}
> 1950$ mJy kpc$^{2}$ for most periods. We assume that the Galactic
and SMC luminosity distributions are similar.  Lorimer et
al.~(1993)\nocite{lbdh93} show that for the Galactic population, the
number of pulsars above a luminosity threshold $L_{0}$ scales as the
inverse of $L_{0}$. Thus, the expected number of pulsars above our
detection limits in the SMC is 0.5 $\pm$ 0.1 non-recycled pulsars and
0.5 $\pm$ 0.2 millisecond pulsars. The number of detectable
millisecond pulsars is an upper limit since this number may be
overestimated for two reasons.  First, our survey sensitivity begins
to significantly decrease at millisecond periods (see Figure
\ref{fig-2}), and second, the younger age of the oldest star clusters
in the SMC relative to the Galaxy (\cite{osm+96a}) may indicate that
the SMC presently contains fewer old recycled pulsars.  In any case,
with two known SMC pulsars, both of which are non-recycled, and
assuming Poisson statistics, the prediction is consistent with our
results.

We can also estimate the number of pulsars detectable in our survey using
the supernova rate for irregular galaxies, of which the SMC is
one. Turatto (2000)\nocite{tur00} estimates that the Type II
supernova rate for irregular galaxies is 0.65 $\pm$ 0.39 SNu, where
SNu is a unit defined as

\begin{equation}
{\rm SNu} = \frac{1}{100~{\rm yr}} \frac{1}{10^{10} L^{B}_{\odot}}
\end{equation} 

\noindent
where $L^{B}_{\odot}$ is the bolometric luminosity of the galaxy in
solar units and where $H_{0} = 75$ km s$^{-1}$ Mpc$^{-1}$ is assumed.

Vangioni-Flam et al.~(1980)\nocite{vlm+80} estimate the SMC
bolometric luminosity to be 7.8 $\times$ $10^{8}$ $L_{\odot}$,
implying a Type II supernova rate of (5 $\pm$ 3)~$\times$~10$^{-4}$
per year. We assume that all compact objects formed in these
supernovae are NSs, and that all NSs formed are active pulsars with
$L_{400} >$ 1 mJy kpc$^{2}$ with a mean lifetime of 10 Myr. This
yields 5100 $\pm$ 3600 active pulsars in the SMC. We apply the
Biggs (1990)\nocite{big90b} beaming fraction of 0.3 for non-recycled
pulsars in the range 0.1 s $<$ $P$ $<$ 1 s to get 1520 $\pm$ 1070
potentially observable pulsars in the SMC. Scaling this to the number
expected to be detectable in our survey using the luminosity law of
Lorimer et al.~(1993)\nocite{lbdh93} gives 0.8 $\pm$ 0.5 for the
predicted number of detectable non-recycled pulsars in our survey.
Our observed sample of two is consistent with this prediction.

\subsection{The Magellanic Cloud Pulsar 
Luminosity Distribution on the $P$-$\dot{P}$ Diagram}

Distances to pulsars in our Galaxy are generally determined by the
DM-distance model of Taylor \& Cordes~(1993)\nocite{tc93} which has
large uncertainties. This unfortunately introduces large uncertainties
in pulsar luminosity estimates. The Magellanic Clouds have more
accurately determined distances than Galactic pulsars, and therefore
their pulsars have better known luminosities. Since only the upper end
of the luminosity function is detectable in the Magellanic Clouds, we
can use luminosity estimates of their pulsars to compare the high end
of the Magellanic Cloud pulsar luminosity function with that of the
Galaxy.

\begin{figure}
\plotone{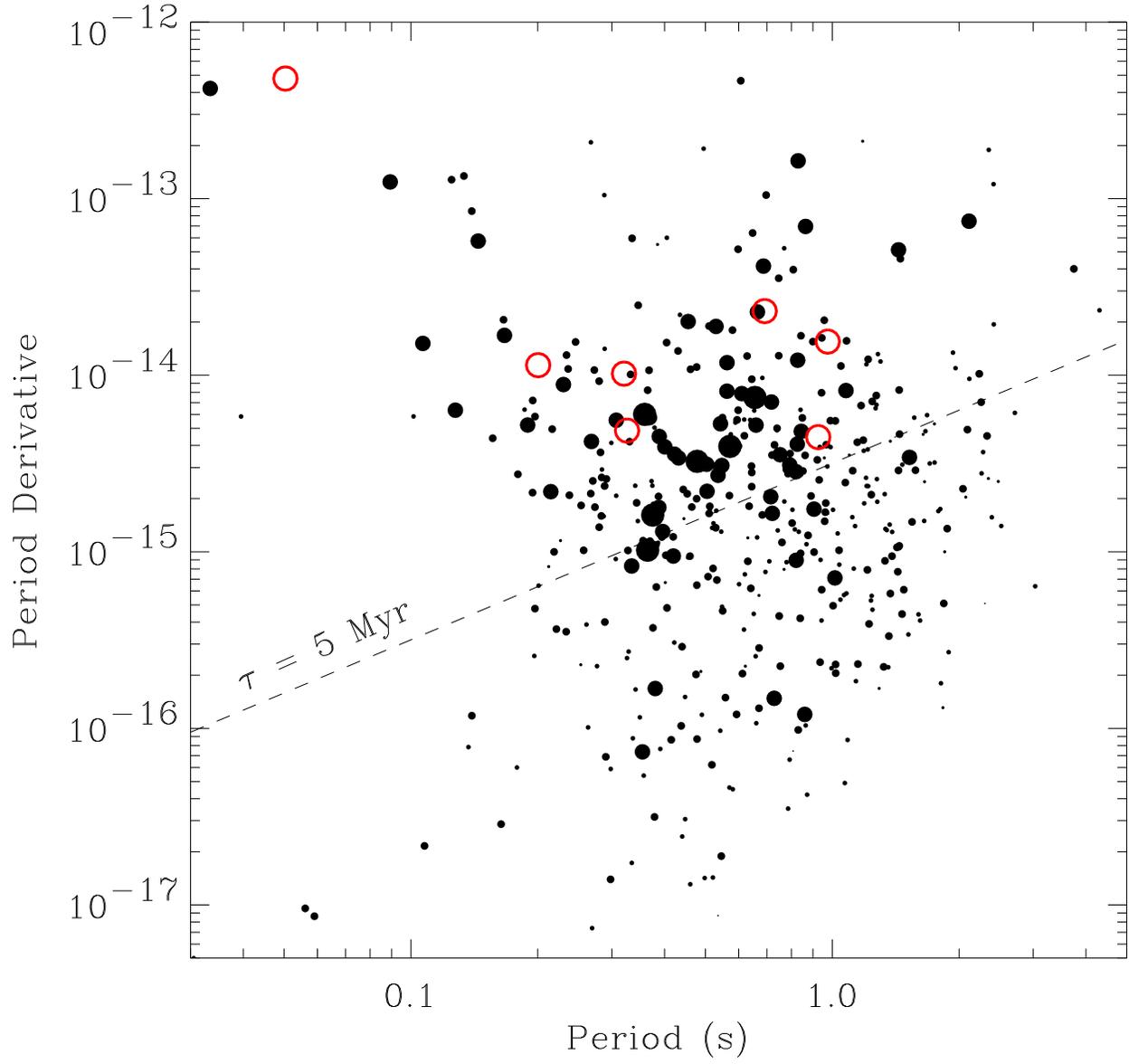}
\caption{$P$-$\dot{P}$ diagram for a subset of Galactic pulsars with
cataloged radio luminosities (filled circles). The symbol size scales
logarithmically with increasing radio luminosity. Also shown are the
seven Magellanic Cloud pulsars for which a radio luminosity has been
estimated (open circles). The dashed line is the 5-Myr isochrone.}
\label{fig-7}
\end{figure}

Table \ref{tbl-4new} lists radio luminosities for the seven known
radio pulsars in the Magellanic Clouds using flux density estimates
from several sources.  Taylor, Manchester, \&
Lyne~(1993)\nocite{tml93} have shown that Galactic pulsars older than
$\sim$ 5 Myr have significantly lower luminosities than younger
pulsars. Figure \ref{fig-7} shows a $P$-$\dot{P}$ diagram for the
seven Magellanic Cloud radio pulsars and Galactic pulsars with
cataloged radio luminosities. The distribution of the Magellanic Cloud
pulsars is consistent with the Galactic distribution of luminous
pulsars and supports the suggestion of Taylor, Manchester, \&
Lyne~(1993)\nocite{tml93}.

\section{Conclusions} 

We have conducted a survey of the SMC for radio pulsars and have
discovered two new pulsars, one of which is located within the SMC. We
present timing results for both of these pulsars as well as refined
values for three previously known LMC pulsars. We have also discovered
serendipitously a new pulsar in the LMC, bringing the total number of
known rotation-powered pulsars in the Magellanic Clouds to eight. The
results of our survey are consistent with the expected number of
detectable SMC pulsars estimated using several methods. However, the
small number of pulsars currently known in the SMC prevents
significant constraints on the parameter assumptions made in these
estimates. The luminosity distribution of the Magellanic Cloud pulsar
population on the $P$-$\dot{P}$ diagram is consistent with the
Galactic distribution of luminous pulsars and supports the conclusion
of Taylor, Manchester, \& Lyne~(1993)\nocite{tml93} that pulsars
younger than $\sim$ 5 Myr are more luminous on average than older
pulsars. The DM distribution of the newly discovered pulsars is
consistent with the known pulsar population in the Magellanic
Clouds. Both of the known SMC pulsars are in a narrow DM range, and if
additional SMC pulsars are found in the future which are also confined
to this narrow range, this will be at odds with the SMC depth
estimated from Cepheid observations. Significantly increasing the
sensitivity of searches for pulsars in the Magellanic Clouds is not
practical with current technology given constraints on telescope time,
etc. However, next-generation instruments such as a square-kilometer
array should greatly increase the number of known pulsars in the
Magellanic Clouds and will enhance and refine the conclusions drawn in
this paper.

\nocite{cra00} 

\acknowledgments

We thank members of the Parkes Multibeam Pulsar Survey team for
assistance with the observations and the dedicated staff at Parkes for
their support during this project. The Parkes radio telescope is part of
the Australia Telescope, which is funded by the Commonwealth of
Australia for operation as a National Facility managed by CSIRO. This
work was supported by an NSF Career Award to VMK (AST-9875897) and
NASA grant NAG~5-3229 to F. Camilo.

\newcommand{\noopsort}[1]{} \newcommand{\printfirst}[2]{#1}
  \newcommand{\singleletter}[1]{#1} \newcommand{\switchargs}[2]{#2#1}

\newpage

\newpage






\begin{deluxetable}{llll}
\footnotesize    
\tablecaption{Astrometric and Spin Parameters for  
Newly Discovered Pulsars\label{tbl-1}}
\tablehead{
\colhead{Name} & 
\colhead{J0057$-$7201} & 
\colhead{J0113$-$7220} &
\colhead{J0535$-$6935}
} 
\startdata
Right Ascension, $\alpha$ (J2000)                                & $00^{\rm h} 57^{\rm m} 44^{\rm s}.0$(4)         & $01^{\rm h} 13^{\rm m} 11^{\rm s}.09$(3)            & $05^{\rm{h}} 35^{\rm{m}}$(2)     \nl
Declination, $\delta$ (J2000)                                    & $-72^{\circ} 01^{\prime} 19^{\prime \prime}$(2) & $-72^{\circ} 20^{\prime} 32^{\prime \prime}.20$(15) & $-69^{\circ} 35^{\prime}$(7)  \nl
Period, $P$ (ms)                                                 & 738.0624426(2)   & 325.88301613(1)        & 200.51011(2) \nl
Period Derivative, $\dot{P}$ ($\times 10^{-15}$)                 & 0.10(8)          & 4.8590(15)             & 11.4(8)\tablenotemark{a} \nl
Dispersion Measure, DM (pc cm$^{-3}$)                            & 27(5)            & 125.49(3)              & 89.4(8)      \nl
Epoch of Period (MJD)                                            & 51213.0          & 51212.0                & 51006.8      \nl
RMS residual (ms)                                                & 1.2              & 0.3                    & --           \nl
Number of TOAs (20cm/50cm/70cm)                                  & 16/0/0           & 35/0/3                 & --           \nl
Timing span (days)                                               & 300              & 430                    & --           \nl
Characteristic age, $\tau_{c}$ (Myr)\tablenotemark{b}            & $\sim$ 100       & 1.1                    & 0.28         \nl
Surface magnetic field $B$ ($\times 10^{12}$ G)\tablenotemark{c} & $\sim$ 0.3       & 1.3                    & 1.5          \nl
Spin-down luminosity, $\dot{E}$ (erg s$^{-1}$)\tablenotemark{d}  & $\sim$ 10$^{31}$ & 5.5 $\times$ 10$^{33}$ & 5.6 $\times$ 10$^{34}$ \nl
Notes                                                            & foreground       & in SMC                 & in LMC, no timing info. \nl
\enddata

\tablenotetext{}{Figures in parentheses represent 1$\sigma$ uncertainties
in the least-significant digit quoted.}

\tablenotetext{a}{Estimated by comparing the barycentric period in
observations taken one year apart.}

\tablenotetext{b}{$\tau_{c} \equiv P/2\dot{P}$.}

\tablenotetext{c}{$B \equiv 3.2 \times 10^{19} (P\dot{P})^{1/2}$ G.}

\tablenotetext{d}{$\dot{E} \equiv 4\pi^{2}I\dot{P}/P^{3}$.}

\end{deluxetable}

\begin{deluxetable}{llll}
\footnotesize  
\tablecaption{Refined Astrometric and Spin Parameters
for Three Previously Known Pulsars\label{tbl-2}}
\tablehead{
\colhead{Name} & 
\colhead{J0455$-$6951} & 
\colhead{J0502$-$6617} & 
\colhead{J0529$-$6652}  
} 
\startdata
Right Ascension, $\alpha$ (J2000)                                & $04^{\rm h} 55^{\rm m} 47^{\rm s}.55$(8) & $05^{\rm h} 02^{\rm m} 50^{\rm s}.53$(10) & $05^{\rm h} 29^{\rm m} 50^{\rm s}.92$(13)  \nl
Declination, $\delta$ (J2000)                                    & $-69^{\circ} 51^{\prime} 34^{\prime \prime}.3$(6) & $-66^{\circ} 17^{\prime} 58^{\prime \prime}.8$(9) & $-66^{\circ} 52^{\prime} 38^{\prime \prime}.2$(9) \nl
Period, $P$ (ms)                                                 & 320.422711526(12)& 691.25141818(4)  & 975.72496638(6)  \nl
Period Derivative, $\dot{P}$ ($\times 10^{-15}$)                 & 10.2119(15)      & 23.090(6)        & 15.509(6)        \nl
Dispersion Measure, DM (pc cm$^{-3}$)                            & 94.89(14)        & 68.9(3)          & 103.2(3)         \nl
Epoch of Period (MJD)                                            & 48757.0          & 48771.0          & 48739.0          \nl
RMS residual (ms)                                                & 1.6              & 2.2              & 3.1              \nl
Number of TOAs (20cm/50cm/70cm)                                  & 1/7/28           & 2/0/27           & 4/1/28           \nl
Timing span (days)                                               & 850              & 850              & 900              \nl
Characteristic age, $\tau_{c}$ (Myr)                             & 0.50             & 0.48             & 1.0              \nl
Surface magnetic field $B$ ($\times 10^{12}$ G)                  & 1.8              & 4.0              & 3.9              \nl
Spin-down luminosity, $\dot{E}$ (erg s$^{-1}$)                   & 1.2 $\times$ 10$^{34}$ & 2.8 $\times$ 10$^{33}$ & 6.6 $\times$ 10$^{32}$ \nl
Notes                                                            & in LMC           & in LMC           & in LMC           \nl
\enddata

\tablenotetext{}{Parameters as in Table \ref{tbl-1}.} 

\end{deluxetable}

\begin{deluxetable}{lccccccccc}
\tiny
\tablecaption{Currently Known Magellanic Cloud Pulsars\label{tbl-4new}}
\tablehead{
\colhead{Name} &
\colhead{$P$} &
\colhead{DM} &
\colhead{Radio/X-ray?} &
\colhead{Timing} & 
\colhead{$S_{1374}$} &
\colhead{$S_{1374}$ Flux} &
\colhead{$S_{610}$} &
\colhead{$S_{610}$ Flux} &
\colhead{$L_{400}$\tablenotemark{a}} \nl 
\colhead{} &
\colhead{(ms)} &
\colhead{(pc cm$^{-3}$)} &
\colhead{} &
\colhead{Ref.} &
\colhead{(mJy)} &
\colhead{Ref.} &
\colhead{(mJy)} &
\colhead{Ref.} &
\colhead{(mJy kpc$^{2}$)}   
}
\startdata
SMC J0045$-$7319\tablenotemark{b}   & 926 & 105 & y/n & 1 & 0.3 $\pm$ 0.1  & 2 & 1.0 $\pm$ 0.2               & 3 & 7000 $\pm$ 2500 \nl
SMC J0113$-$7220                    & 326 & 125 & y/n & 2 & 0.4 $\pm$ 0.1  & 2 & --                          &-- & 9500 $\pm$ 2500 \nl
LMC J0535$-$6935\tablenotemark{c}   & 201 & 89  & y/n & 2 & $\sim$ 0.05    & 2 & --                          &-- & $\sim$ 1000     \nl
LMC J0537$-$6910\tablenotemark{c,d} &  16 & $-$ & n/y & 4 & $<$ 0.06       & 5 & $<$ 0.2\tablenotemark{f}    & 5 & $<$ 1200        \nl
LMC J0540$-$6919\tablenotemark{c,e} &  50 & 146 & y/y & 6 &     --         &-- & $\sim$ 0.4\tablenotemark{g} & 6 & $\sim$ 2000     \nl
LMC J0455$-$6951                    & 320 & 95  & y/n & 2 &     --         &-- & 1.0 $\pm$ 0.5               & 3 & 4900 $\pm$ 2500 \nl
LMC J0502$-$6617                    & 691 & 69  & y/n & 2 &     --         &-- & 0.7 $\pm$ 0.4               & 3 & 3500 $\pm$ 2000 \nl
LMC J0529$-$6652                    & 976 & 103 & y/n & 2 & 0.3 $\pm$ 0.1  & 2 & 1.8 $\pm$ 0.8               & 3 & 6000 $\pm$ 1500 \nl
\enddata

\tablenotetext{}{(1) Kaspi et al.~(1994)\nocite{kjb+94}}
\tablenotetext{}{(2) this work} 
\tablenotetext{}{(3) McConnell et al.~(1991)\nocite{mmh+91}}
\tablenotetext{}{(4) Marshall et al.~(1998)\nocite{mgz+98}}
\tablenotetext{}{(5) Crawford et al.~(1998)\nocite{ckm+98a}}
\tablenotetext{}{(6) Manchester et al.~(1993)\nocite{mml+93}}

\tablenotetext{a}{Distances to the LMC and SMC are assumed to be 50
kpc and 57 kpc respectively (\cite{col98}).}
\tablenotetext{b}{Binary with B star companion.}
\tablenotetext{c}{Located in the 30 Doradus region of the LMC.} 
\tablenotetext{d}{Fastest known non-recycled pulsar, has associated plerion.} 
\tablenotetext{e}{Crab twin, has associated plerion.} 
\tablenotetext{f}{660 MHz flux density estimate.} 
\tablenotetext{g}{640 MHz flux density estimate.}

\end{deluxetable}

\end{document}